\documentstyle[prd,aps,eqsecnum,preprint,tighten]{revtex}
\draft
\def\beqa{\begin{eqnarray}}
\def\eeqa{\end{eqnarray}}
\def\beq{\begin{equation}}
\def\eeq{\end{equation}}

\begin{document}
\title{Casimir Energy of a Ball   \\
       and Cylinder in the Zeta Function Technique }
\author{G. Lambiase\thanks{E-mail: lambiase@vaxsa.csied.unisa.it,
lambiase@physics.unisa.it}}
\address{Dipartimento di Scienze Fisiche "E.R. Caianiello" \\
Universit\'a di Salerno, 84081 - Baronissi (SA), Italy \\
Istituto Nazionale di Fisica Nucleare, Sez. Napoli, Italy}
\author{V.V. Nesterenko\thanks{E-mail: nestr@thsun1.jinr.ru}}
\address{Bogoliubov Laboratory of Theoretical Physics \\
Joint Institute for Nuclear Research, Dubna, 141980, Russia}
\author{M. Bordag\thanks{E-mail: Michael.Bordag@itp.uni-leipzig.de}}
\address{Universit\"at Leipzig, Institute f\"ur Theoretical Physik \\
Augustusplatz 10, 04109 Leipzig, Germany}
\date{\today}
\maketitle
\begin{abstract}
A simple method is proposed to construct the spectral zeta
functions required for calculating the electromagnetic vacuum energy
with boundary conditions
given on a sphere or on an infinite cylinder.
When calculating the Casimir energy in this approach no exact
divergencies appear and no renormalization is needed.
The starting point
of the consideration is the representation of the zeta functions in terms
of contour integral, further the uniform asymptotic expansion of
the Bessel function is essentially used. After the analytic
continuation, needed for calculating the Casimir energy, the zeta
functions are presented as infinite series containing the Riemann
zeta function with rapidly falling down terms. The spectral zeta
functions
are constructed exactly for a material ball and infinite cylinder
placed in an uniform endless medium under the condition that the
velocity of light does not change when crossing the
interface. As a special case, perfectly conducting spherical and
cylindrical shells are also considered in the same line. In this
approach one succeeds, specifically, in justifying, in mathematically
rigorous way, the appearance of the contribution to the Casimir
energy for cylinder which is
proportional to $\ln (2\pi)$.
\end{abstract}
\thispagestyle{empty}

\section{Introduction}

A considerable achievement in theoretical investigations of the
Casimir effect~\cite{PGM,MT} was its calculation for massive fields
(scalar and spinor) with boundary conditions on a sphere
\cite{BEKL,ebk98}.  The various divergent contributions had been
discussed in detail from the point of view of the general theory of
adiabatic expansions (resp. heat kernel expansion). In a subsequent
paper \cite{BKV} it was clarified in which cases the calculation of
the Casimir energy, after the proper renormalization, yields a
meaningful (unique) result and in which not independently of the
regularization used. For a massive field a well defined result can be
obtained in any case using the normalization condition proposed
there. Instead, for a massless field the heat kernel coefficient $a_2$
must vanish in order to allow for a meaningful calculation of the
Casimir energy. For instance, this is the case for a material body
characterized by a polarizability and a permittivity when the speeds
of light inside and outside are the same or their difference is small.
The vanishing of $a_2$ for the Dirichlet and Robin boundary conditions
(and as a consequence for the conductor and bag boundary conditions) when
taking inside and outside contributions together made
Boyer's~\cite{Boyer}
 and all
subsequent calculations possible and meaningful. When using a clever
regularization (like the zetafunctional one~\cite{EO,Elizalde})
it is even possible to
avoid the appearance of divergencies other than that in the Minkowski
space contribution at all.

Practically every problem on calculation of the Casimir energy
(or force) has been considered
multiply with employment of more and more effective and elaborated
mathematical methods. For
example, the first calculation of the Casimir energy of a perfectly
conducting spherical shell
carried out by T.H.~Boyer in 1968~\cite{Boyer} has required
computer calculations during 3 years
\cite{Milton}. Later this problem was considered in many
papers \cite{Davies,BD,MRS}.
By making use of the modern methods \cite{NP} it can
be solved practically without numerical
calculations (with a
precision of a few percent). It requires only the
 application of the uniform asymptotic
expansion for the Bessel functions.

In recent papers \cite{BNP,MNN} the Casimir energy of a compact
ball~\cite{foot1} and infinite cylinder
has been calculated by making use of the mode--by--mode summation technique. In
these problems two sums appear, over the roots of radial frequency
equation at fixed
value of angular momentum and
then over angular momentum. The either of these sums is divergent.
In papers~\cite{BNP,MNN}
for each of these summation a separate regularization has been used.
The first summation was
carried out by applying the contour integration with subsequent subtraction
of the contribution
of an infinite homogeneous space. The second sum was evaluated by making
use of the Riemann
$\zeta$ function technique. However the procedure of analytic continuation,
required
by rigorous approach, has not been considered there.

The present paper pursues the aim to eliminate the minor points of
 preceding considerations,
i.e., the Casimir energy for two configurations mentioned above will be
calculated by the
rigorous $\zeta$ function techniques, and the analytic continuation
of the relevant spectral $\zeta$ functions
will be carried out exactly. An essential advantage of this
regularization procedure is that
no manifestly divergent expressions arise in its framework,
and it gives final finite result without any subtractions (renormalizations).

The layout of this paper is as follows. In Sec.\ II, the spectral
zeta function is constructed for a compact ball placed in uniform
endless medium when the light velocity is the same inside and
outside the ball. As a special case the zeta function for
perfectly conducting spherical shell is also considered. In Sect.\
III the spectral functions for infinite cylinder are constructed
under the same conditions. These results provide a firm footing
for the previous calculations of the Casimir energy for given
boundary conditions by making use of a "naive" zeta function
method. In Sect.\ IV the obtained results are shortly discussed.

\section{Casimir energy of a compact ball under
the condition $\varepsilon\mu=1$}
In the $\zeta$ function method \cite{EO,Elizalde}
 the Casimir energy $E_C$ is defined in
the following way. Let $\omega_p$'s be the eigenfrequencies of the quantum
field system under the influence of the boundary conditions, and let
$\bar{\omega}_p$'s be the same frequencies when the boundaries are removed.
By making use of this spectrum one defines the $\zeta$ function for the problem
in hand
\beq\label{zeta}
\zeta (s)=\sum_{\{p\}}(\omega_p^{-s}-\bar{\omega}_p^{-s})\,{.}
\eeq
Here the summation (or integration) should be done over all the quantum
numbers $\{p\}$ specifying the spectrum. The parameter $s$ is considered
at first to belong to region of the complex plane $s$ where the sum
(\ref{zeta}) exists. Further the analytic continuation of
(\ref{zeta}) to the point $s=-1$ should be constructed. After that one puts
\beq\label{E_C}
E_C=\frac{1}{2}\zeta(s=-1)\,{.}
\eeq

Let us consider a solid ball of radius $a$, consisting of a material which
is characterized by permittivity $\varepsilon_1$ and permeability $\mu_1$.
The ball is assumed to be placed in an infinite medium with permittivity
$\varepsilon_2$ and permeability $\mu_2$.
The eigenfrequencies of electromagnetic field for this configuration are
determined by the frequency equation for the TE--modes \cite{Str}
\beq\label{TE}
\Delta_l^{\text{TE}}(a\omega)\equiv \sqrt{\varepsilon_1\mu_2}
\,\tilde{s}^{\prime}_l
(k_1a)\,\tilde{e}_l(k_2a)-
\sqrt{\varepsilon_2\mu_1}\,\tilde{s}_l(k_1a)\,\tilde{e}^{\prime}_l(k_2a)=0\,{,}
\eeq
and the analogous equation for the TM--modes
\beq\label{TM}
\Delta_l^{\text{TM}}(a\omega)\equiv \sqrt{\varepsilon_2\mu_1}
\,\tilde{s}^{\prime}_l
(k_1a)\,\tilde{e}_l(k_2a)-
\sqrt{\varepsilon_1\mu_2}\,\tilde{s}_l(k_1a)\,\tilde{e}^{\prime}_l(k_2a)=0\,{,}
\eeq
where $k_i=\sqrt{\varepsilon_i\mu_i}\,\omega, i=1,2$ are the wave numbers
inside and outside the ball, respectively. Here $\tilde{s}_l(x)$ and
$\tilde{e}_l(x)$ are the Riccati--Bessel functions
\beq\label{set}
\tilde{s}_l(x)=\sqrt{\frac{\pi}{2x}}\, J_{l+1/2}(x)\,{,} \qquad
\tilde{e}_l(x)=\sqrt{\frac{\pi}{2x}}\,H^{(1)}_{l+1/2}(x)\,{,}
\eeq
and prime stands for the differentiation with respect to their arguments, $k_1a$
or $k_2a$. The orbital momentum $l$ in Eqs. (\ref{TE}) and (\ref{TM})
assumes the values $1,2, \ldots$.

As usual when one is dealing with an analytic continuation, it is convenient
to represent the sum (\ref{zeta}) in terms of the contour integral
\beq\label{zeta_C}
\zeta_C(s)=\frac{2l+1}{2\pi i}\lim_{\mu\to 0}\oint_C dz (z^2+\mu^2)^{-s/2}
\frac{d}{dz}\ln
\frac{\Delta_l^{\text{TE}}(az)\Delta_l^{\text{TM}}
(az)}{\Delta_l^{\text{TE}}(\infty)
\Delta_l^{\text{TM}}
(\infty)}\,{,}
\eeq
where the contour $C$ surrounds, counterclockwise, the roots of the
frequency equations in the right half--plane. For brevity we write in
(\ref{zeta_C}) simply $\Delta_l(\infty)$
instead of $\lim_{a\to\infty}\Delta_l(az)$. Transition to the complex
frequencies $z$ in Eq. (\ref{zeta_C}) is accomplished by introducing the
unphysical photon mass $\mu$
\beq\label{mu}
\omega \rightarrow (z^2+\mu^2)^{s/2}\vert_{\mu\to 0}\,{.}
\eeq
Extension to the complex $z$--plane of the frequency equations
$\Delta_l^{\text{TE}}(az)$
and $\Delta_l^{\text{TM}}(az)$ should be done in the following way. In the
upper (lower)
half--plane the Hankel functions of the first (second) kind $H_{\nu}^{(2)}(az)$
($H_{\nu}^{(1)}(az)$) must be used~\cite{foot2}.
Location of the roots of Eqs.
(\ref{TE}) and (\ref{TM}) enables one to deform the contour $C$ into a
segment of the imaginary axis
$(-i\Lambda, i\Lambda)$ and a semicircle of radius $\Lambda$ in the right
half--plane. When $\Lambda$
tends to infinity the contribution along the semicircle into $\zeta_{ball} (s)$
vanishes because the argument of the logarithmic function in the integrand
tends in this case to 1.
As a result we obtain
\beq\label{zetaim}
\zeta_{ball}(s)=-\frac{(2l+1)a^s}{2\pi i}\lim_{\mu\to 0}
\int_{-i\infty}^{+i\infty}dz
(z^2+\mu^2)^{-s/2}
\frac{d}{dz}\ln\frac{\Delta_l^{\text{TE}}(z)\Delta_l^{\text{TM}}(z)}
{\Delta_l^{\text{TE}}(i\infty)
\Delta_l^{\text{TM}}(i\infty)}\,{.}
\eeq
Now we impose the condition that the velocity of light inside and outside
the ball is the same
\beq\label{vel}
\varepsilon_1\mu_1=\varepsilon_2\mu_2=c^{-2}\,{.}
\eeq
Under this assumption the argument of the
logarithm in (\ref{zetaim}) can be simplified
considerably \cite{BNP} with the result
\beq\label{zetareal}
\zeta_{ball}(s)=\left(\frac{c}{a}\right)^{-s}\sum_{l=1}^{\infty}(2l+1)\,
\frac{\sin(\pi s/2)}{\pi}\int_0^{\infty}dy\, y^{-s}\frac{d}{dy}
\ln[1-\xi^2\sigma_l^2(y)]\,{,}
\eeq
where
\beq\label{xisigma}
\xi=\frac{\varepsilon_1-\varepsilon_2}{\varepsilon_1+\varepsilon_2}, \quad
\sigma_l (y)=\frac{d}{dy}[s_l(y)\,e_l(y)]\,{.}
\eeq
Here $s_l(y)$ and $e_l(y)$ are the modified Riccati--Bessel functions
\beq\label{se}
s_l(x)=\sqrt{\frac{\pi x}{2}}I_{\nu}(x),\quad
e_l(x)=\sqrt{\frac{2 x}{\pi}}K_{\nu}(x),
\quad \nu=l+1/2\,{.}
\eeq
More details concerning the contour integral representation of the spectral
$\zeta$ functions can be found in \cite{BEKL,Bordag,ELR,LR}.

Further the analytic continuation of Eq.\ (\ref{zetareal}) is accomplished
by expressing the
sum over $l$ in terms of the Riemann $\zeta$ function. This cannot be done
in a closed form.
Making use of the uniform asymptotic expansion (UAE)  for the Bessel
functions in increase
powers of $1/\nu$ enables one to construct the analytic continuation looked
for in the form of
the series, the terms of which are expressed through the Riemann $\zeta$
function.
The problem of the convergence of this series does not arise because its
terms go down very
fast.

We demonstrate this keeping only two terms in UAE for the product of the
Bessel functions
$I_{\nu}(\nu z)K_{\nu}(\nu z)$ \cite{AS}
\beq\label{UAE}
I_{\nu}(\nu z)K_{\nu}(\nu z)\simeq \frac{t}{2\nu}
 \left[1+\frac{t^2(1-6t^2+t^4)}{8\nu^2}+\ldots
\right],\quad t=\frac{1}{\sqrt{1+z^2}}\,{.}
\eeq
After changing the integration variable $y=\nu z$ in
Eq. (\ref{zetareal}) we substitute
(\ref{UAE}) into this formula and expand the
logarithm function up to the order $\nu^{-3}$
keeping at the same time only the terms linear in $\xi^2$.
 The last assumption is not principal.
It is introduced for simplicity and in order
to have possibility of a direct comparison with
the results of Ref.~\cite{BNP}. Thus we have
\beq\label{exp}
\frac{d}{dz}\ln\left\{1-\xi^2\left[\frac{d}{dz}(zI_{\nu}(\nu z)K_{\nu}
(\nu z))\right]^2\right\}
=
\eeq
$$
=\frac{3}{2}\frac{\xi^2}{\nu^2}\,zt^8+\frac{\xi^2}{16\nu^4}\,zt^8(-12+216t^2
-600t^4+420t^6)
+O(\nu^{-6})\,{.}
$$
Integration over $z$ can be done by making use of the formula
\beq\label{beta}
\int_0^{\infty}z^{-\alpha-1}t^{\beta}dz=\frac{1}{2}
\frac{\Gamma \left(\displaystyle \frac{\alpha+\beta}{2}\right)
\Gamma \left(-\displaystyle \frac{\alpha}{2}\right)}{\Gamma \left(\displaystyle
\frac{\beta}{2}\right)}\,{.}
\eeq
Also the properties of the $\Gamma$ function
\beq\label{gamma}
\Gamma (z)\Gamma (1-z)=\frac{\pi}{\sin\pi z}, \qquad \Gamma (1+z)=z\Gamma (z)
\eeq
prove to be useful. After simple calculations we arrive at the result
\beq\label{in}
\zeta_{ball}(s)\simeq
\frac{\xi^2}{32}\left(\frac{c}{a}\right)^{-s}s(2+s)(4+s)\left(
\sum_{l=1}^{\infty}\nu^{-1-s}+p(s)\sum_{l=1}^{\infty}\nu^{-3-s}+\ldots
\right)\,{,}
\eeq
$$
\nu=l+1/2\,{,}
$$
where
\beq\label{pol}
p(s)=-\frac{1}{2}\left[1-\frac{9}{4}(6+s)
+\frac{5}{8}(6+s)(8+s)-\frac{7}{192}(6+s)(8+s)(10+s)
\right]\,{.}
\eeq
The zeta function $\zeta_{ball}(s)$ represented in the form (\ref{in}) is
defined for Re~$s> 0$ due
to the first sum over $l$. This term corresponds to the order $1/\nu$ in
the uniform asymptotic
expansion (\ref{UAE}). The second sum in (\ref{in}), defined at Re~$s>-2$,
has been generated by
the term $\sim 1/\nu^3$ in Eq.\ (\ref{UAE}). It is clear that
the terms of order $1/\nu^{2k+1}$ in (\ref{UAE}) will give rise
 to the singularity of
$\zeta_{ball}(s)$ at the points $s=-2k,\quad k=0,1,2,\ldots$. Due to
the multipliers in front of the
square brackets in (\ref{in}) the first three singularities are really the
indefinitenesses like
$0\cdot \infty$.

The analytic continuation of Eqs.~(\ref{in}), (\ref{pol}) into the region
Re~$s\leq 0$
can be accomplished by expressing the sums over angular momentum $l$ trough
the Riemann
$\zeta$ function according to the formula \cite{GR}
\beq\label{hur}
\sum_{l=1}^{\infty}\nu^{-s}=(2^s-1)\zeta (s)-2^s\,{,}\qquad \nu=l+1/2\,{.}
\eeq
As a result one gets
\beq\label{con}
\zeta_{ball}(s)\simeq \frac{\xi^2}{32}\left(\frac{c}{a}
\right)^{-s}s(2+s)(4+s)\{(2^{1+s}-1)
\zeta(1+s)-2^{1+s}
\eeq
$$
+p(s)[(2^{3+s}-1)\zeta(1+s)-2^{3+s}]+\ldots \}\,{.}
$$
The singularities in Eq. (\ref{in}) are transformed
in (\ref{con}) into the poles of the
Riemann $\zeta$ functions at the points $s=2k, \quad k=0,1,2,\ldots$
\beqa
\zeta(1+s) & \simeq & \displaystyle \frac{1}{s}+\gamma+\ldots, \quad s\to 0
\,{,} \nonumber \\
\zeta(3+s) & \simeq & \displaystyle\frac{1}{s+2}+\gamma+\ldots,\quad s\to -2
\,{,}\label{poles} \\
\zeta(5+s) & \simeq & \displaystyle\frac{1}{s+4}+\gamma+\ldots, \quad s\to -4
\,{,} \nonumber \\
.........  & ...... & ............................... {\,},\nonumber
\eeqa
where $\gamma$ is the Euler constant. The first three poles are annihilated
by the multipliers
in front of the curly brackets in Eq.\ (\ref{con}). The first surviving
singularity
(simple pole) appears only at the point $s=-6$.
Thus the formula (\ref{con}) affords the required analytic continuation of the
function $\zeta_{ball}(s)$ into the region Re~$s< 0$. In view of Eq.\
 (\ref{E_C}) we are
interested in the point $s=-1$ where $\zeta_{ball}(s)$ given by (\ref{con})
is regular
\beq\label{final}
\zeta_{ball}(-1)=\frac{3\xi^2c}{32a}\left[1+\frac{9}{128}\left(\frac{\pi^2}{2}
-4\right)+
\ldots \right]\,{.}
\eeq
It is exactly the first two terms in Eq. (3.10) of Ref.~\cite{BNP}. The
procedure of
analytic continuation presented above can be extended in a straightforward
way to the arbitrary
order of the uniform asymptotic expansion (\ref{UAE}). Certainly in this case
analytical
calculations should be done by making use of {\it Mathematica} or {\it Maple}.

The problem under consideration with $\xi=1$ is of a special interest because
in this case it
gives the Casimir energy of a perfectly conducting spherical shell.
As it was noted above, this
configuration has been considered by many authors. We present here the basic
formulae which
afford the analytical continuation of the corresponding spectral
$\zeta$ function.
We again content ourselves two terms in the UAE (\ref{UAE}).
In the next formula (\ref{exp})
It is impossible to put simply $\xi=1$ in the next formula (\ref{exp}).
One has to do the expansion here anew keeping all the terms
$\sim 1/\nu^4$. This gives
\beq\label{shell1}
\frac{d}{dz}\ln\left\{1-\left[\frac{d}{dz}(zI_{\nu}(\nu z)K_{\nu}(\nu
z))\right]^2\right\}=
\eeq
$$
=
\left[\frac{3}{2\nu^2}zt^8+\frac{3}{4\nu^4}zt^8
\left(-1+18t^2-50t^4+35t^6\right)+
O(\nu^{-6})\right]\,{.}
$$
After integration and elementary simplifications
we arrive at the following result
for the spectral function in hand
\beq\label{shell2}
\zeta_{shell}(s)\simeq \frac{1}{32a^{-s}}\, s(2+s)(4+s)
\left[\sum_{l=1}^{\infty}\nu^{-1-s}+
q(s)\sum_{l=1}^{\infty}\nu^{-3-s}+\ldots \right]\,{,}
\eeq
where
\beq\label{qpol}
q(s)= \frac{1}{3840}\,(480+868s+504s^2+71s^3)\,{.}
\eeq
Obviously formula (\ref{shell1}) has the same singularities as
Eq. (\ref{in}), i.e., it is
defined for Re~$s> 0$. The analytic continuation is
accomplished by making use of
Eq.~(\ref{hur})
\beq\label{shell3}
\zeta_{shell}(s)\simeq\frac{1}{32a^{-s}}\, s(2+s)(4+s)\{(2^{1+s}-
1)\zeta (1+s)-2^{1+s}+
\eeq
$$
+q(s)[(2^{3+s}-1)\zeta (3+s)-2^{3+s}]+\ldots\}\,{.}
$$
The nearest singularity in this formula is simple pole at $s=-6$. As above it is
originated in the term $\sim 1/\nu^7$ in the UAE (\ref{UAE}).
At the point $s=-1$
the spectral zeta function $\zeta_{shell}(s)$ is regular and gives
the following value for the Casimir energy of a perfectly
conducting spherical shell
\beq\label{shell4}
E_{shell}(-1)=\frac{1}{2}\,\zeta_{shell}(-1)=\frac{3}{64a}\left[1-\frac{3}{256}
\left(\frac{\pi^2}{2}-4\right)+\ldots \right]=\frac{1}{a}\, 0.046361\ldots \,{.}
\eeq
Without considering the analytic continuation and do not carrying
out the analysis of the
singularities in the complex $s$ plane this result has been
obtained in \cite{NP}.

Undoubtedly, the calculation of the Casimir energy of a
nonmagnetic dielectric ball
($\varepsilon_1\mu_1\neq \varepsilon_2\mu_2$) by a rigorous $\zeta$
function method is also of
a special interest. However, in this case the very definition of
the spectral zeta function
should probably be changed in order to incorporate  the contact
terms which seem to be essential
in this problem \cite{MNg,BM,BMM}.

\section{Vacuum energy of electromagnetic
field with boundary conditions given on an infinite
cylinder}
Calculation of the Casimir energy of an infinite cylinder~\cite{RM,MNN}
proves to be a more
involved problem in comparison with that for sphere.
In this section the spectral zeta
function $\zeta_{cyl}(s)$, for this configuration will
be constructed its analytical
continuation into the left half--plane of the complex variable $s$ will be
carried out, and relevant singularities
will be analyzed.

Thus we are considering an infinite cylinder of radius
$a$ which is placed in an uniform
unbounded medium. The permittivity and the permeability of the material
making up the cylinder
are $\varepsilon_1$ and $\mu_1$, respectively,
and those for surrounding medium are
$\varepsilon_2$ and $\mu_2$. We assume again that
the condition (\ref{vel}) is fulfilled.
In this case the electromagnetic oscillations can again
be divided into the transverse--electric (TE)
modes and transverse--magnetic (TM) modes. In terms of the
cylindrical coordinates
$(r, \theta, z)$ the eigenfunctions of the given boundary value problem
contain the multiplier
\beq\label{mult}
\exp{(\pm i\omega t+ik_zz+in\theta)}
\eeq
and their dependence on $r$ is described by the Bessel
functions $J_n$ for $r< a$
and by the Hankel functions $H_n^{(1)}$ or $H_n^{(2)}$ for $r> a$.
The frequencies of TE-- and TM--modes are determined, respectively,
 by the equations \cite{Str}
\beqa
\Delta_n^{\text{TE}}(\lambda , a)\equiv & \lambda
a[\mu_1J_n^{\prime}(\lambda a)H_n(\lambda a)-
\mu_1J_n(\lambda a)H_n(\lambda a)^{\prime}]=0 \label{fr1}\,{,} \\
\Delta_n^{\text{TM}}(\lambda , a)\equiv
& \lambda a[\varepsilon_1J_n^{\prime}(\lambda a)H_n(\lambda a)-
\varepsilon_1J_n(\lambda a)H_n(\lambda a)^{\prime}]=0 \label{fr2}\,{,}
\eeqa
where $n=0,\pm 1,\pm 2, \ldots $. Here $\lambda$ is the eigenvalue
of the corresponding
transverse (membrane--like) boundary value problem
\beq\label{lambda}
\lambda^2=\frac{\omega^2}{c^2}-k_z^2\,{.}
\eeq
In a complete analogy with the preceding Section we define the
Casimir energy per unit length of
the cylinder trough the spectral zeta function
\beq\label{E_CYL}
E_{cyl}=\frac{1}{2}\zeta_{cyl}(-1)\,{.}
\eeq
Let $\lambda_{nr}$ be the roots of the frequency equations
(\ref{fr1}) and (\ref{fr2}), then
the function $\zeta_{cyl} (s)$ is introduced in the following way
\beq\label{zcyl1}
\zeta_{cyl}(s)=c^{-s}\int_{-\infty}^{+\infty}\frac{dk_z}{2\pi}\sum_{n,r}
[(\lambda_{nr}(a)+k_z^2)^{-s/2}-(\lambda_{nr}(\infty)+k_z^2)^{-s/2}]\,{.}
\eeq
In terms of the contour integral it can be represented in the form
\beq\label{zcyl2}
\zeta_{cyl}(s)=c^{-s}\int_{-\infty}^{+\infty}\frac{dk_z}{2\pi}
\sum_{n=-\infty}^{+\infty}
\oint_C (\lambda^2+k_z^2)^{-s/2}d_{\lambda}\ln\frac{\Delta_n^{\text{TE}}
(\lambda a)\Delta_n^{\text{TM}}
(\lambda a)}{\Delta_n^{\text{TE}}(\infty)\Delta_n^{\text{TM}}(\infty)}\,{.}
\eeq
Again we can take the contour $C$ to consist of the imaginary
axis $(+i\infty , -i\infty)$
closed by a semicircle of an infinitely large radius in the right half--plane.
Continuation of the expressions $\Delta_n^{\text{TE}}(\lambda a)$
and $\Delta_n^{\text{TM}}(\lambda a)$
into the complex plane $\lambda$ should be done in the same way as in the
preceding Section, i.e.,
by using $H_n^{(1)}(\lambda)$ for Im~$\lambda
< 0$ and $H_n^{(2)}(\lambda)$ for Im~$\lambda > 0$.
On the semicircle the argument of the logarithm in Eq. (\ref{zcyl2})
tends to 1. As
a result this part of the contour $C$ does not give any contribution
into the zeta function
$\zeta_{cyl} (s)$. When integrating along the imaginary
axis we choose the branch line of the
function $\phi(\lambda )=(\lambda^2+k_z^2)^{-s/2}$ to run
between $-ik_z$ and $+ik_z$, where
$k_z=+\sqrt{k_z^2}>0$ and use further that branch
of this function which assumes real values
when $\vert y\vert < k_z$, where $y=$ Im~$\lambda$. More precisely we have
\beq
\phi (iy)=\left\{ \begin{array}{ll}
              e^{-i\pi s/2}(y^2-k_z^2)^{-s/2},  & y>k_z, \nonumber \\
  (k_z^2-y^2)^{-s/2},               & \vert y\vert <k_z,\label{phi} \\
              e^{i\pi s/2}(y^2-k_z^2)^{-s/2},   & y<-k_z.\nonumber
              \end{array}  \right.
\eeq
Employment of the Hankel functions $H_n^{(1)}(\lambda)$
and $H_n^{(2)}(\lambda)$ by extending
the expressions $\Delta_n^{\text{TE}}(\lambda )$
and $\Delta_n^{\text{TM}}(\lambda )$
into the complex
plane $\lambda$, as it was noted above, gives rise to the argument
of the logarithm function
depending only on $y^2$ on the imaginary axis.
It means that the derivative of the logarithm is odd
function of $y$. As a result the segment of the
imaginary axis $(-ik_z, +ik_z)$ gives zero, and
Eq.\ (\ref{zcyl2}) acquires the form
\beq\label{3.9}
\zeta_{cyl}(s)=\frac{c^{-s}}{\pi^2}\sin\frac{\pi
s}{2}\sum_{n=-\infty}^{+\infty}\int_0^{\infty}
dk_z\int_{k_z}^{\infty}(y^2-k_z^2)^{-s/2}d_y
\ln\frac{\Delta^{\text{TE}}(ay)_n\Delta^{\text{TM}}_n(ay)}
{\Delta^{\text{TE}}_n(i\infty)\Delta^{\text{TM}}_n(i\infty)}\,{.}
\eeq
Changing the order of integration of $k_z$ and $y$ and taking into
account the value of the integral
\beq\label{3.10}
\int_0^{y}dk_z(y^2-k_z^2)^{-s/2}=\frac{\sqrt{\pi}}{2}\, y^{1-s}\,
\frac{\Gamma\displaystyle
\left(1-\frac{s}{2}\right)}{\Gamma\left(\displaystyle\frac{3-s}{2}
\right)}\,{,}\qquad
\mbox{Re}~s<2\,{,}
\eeq
we obtain after the substitution $ay\to y$
\beq\label{3.11}
\zeta_{cyl}(s)=\frac{1}{2\sqrt{\pi}\,
a\Gamma\left(\displaystyle\frac{s}{2}\right)\Gamma
\left(\displaystyle\frac{3-s}{2}\right)}
\left(\frac{c}{s}\right)^{-s}\sum_{n=-\infty}^{+\infty}\int_0^{\infty} dy\,
y^{1-s}\frac{d}{dy}\ln[1-\xi^2\mu_n^2(y)]\,{,}
\eeq
where
\beq\label{3.12}
\mu_n(y)=y(I_n(y)K_n(y))^{'}, \qquad
\xi=\frac{\varepsilon_1-\varepsilon_2}{\varepsilon_1+\varepsilon_2}\,{.}
\eeq
We shall again content ourselves with the first two terms in the
uniform asymptotic expansion (\ref{UAE}) and take into account
only the terms linear in $\xi^2$. In this approximation, upon changing
the integration variable $y=nz, n=\pm1, \pm2, \ldots$, we have
\beq\label{3.13}
\ln\left\{1-\xi^2\left[z\frac{d}{dz}(I_n(nz)K_n(nz))\right]^2\right\}=
\eeq
$$
=
-\xi^2\,\frac{z^4t^6}{4n^2}\left[1+\frac{t^2}{4n^2}(3-30t^2+35t^4)
+O(n^{-4})\right]\,{.}
$$
Now we substitute (\ref{3.13}) into all the terms in (\ref{3.11}) with
$n\neq 0$. The term with $n=0$ in this sum will be treated by
subtracting and adding to the logarithmic function the quantity
\beq\label{3.14}
-\frac{\xi^2}{4}\frac{y^4}{(1+y^2)^3}\,{.}
\eeq
As a result the zeta function $\zeta_{cyl}(s)$ can be presented
now as the sum of three terms
\beq\label{3.15}
\zeta_{cyl}(s)=Z_1(s)+Z_2(s)+Z_3(s)\,{,}
\eeq
where
\beqa
Z_1(s)&=& \displaystyle \frac{\displaystyle
\left(\frac{c}{a}\right)^{-s}}{2\sqrt{\pi}a
\Gamma\left(\displaystyle\frac{s}{2}\right)
\Gamma\left(\displaystyle\frac{3-s}{2}\right)}
\displaystyle\int_0^{\infty}dy\,y^{1-s}
\frac{d}{dy}\left\{\ln [1-\xi^2\mu_0^2(y)]+
\frac{\xi^2}{4}\,y^4t^6\right\}, \label{3.16} \\
Z_2(s)&=& -\xi^2\displaystyle\left(\frac{c}{a}
\right)^{-s}\frac{\displaystyle 2\sum_{n=1}^{+\infty}n^{-s-1} +1}
{\displaystyle 8\sqrt{\pi}\,a\,\Gamma
\left(\displaystyle\frac{s}{2}\right)\Gamma\left(\displaystyle\frac{3-s}{2}
\right)} \displaystyle\int_0^{\infty}dz\,z^{1-s}\frac{d}{dz}
 (z^4t^6) \,{,} \label{3.17} \\
Z_3(s)&=& -\xi^2\frac{\displaystyle
\left(\frac{c}{a}\right)^{-s}\displaystyle \sum_{n=1}^{+\infty}n^{-3-s}}
{\displaystyle 32\sqrt{\pi}a\,
\Gamma\left(\displaystyle\frac{s}{2}\right)
\Gamma\left(\displaystyle\frac{3-s}{2}
\right)}\displaystyle\int_0^{\infty}dz\,z^{1-s}
\frac{d}{dz}[z^4t^8(3-30t^2+35t^4)]\,{.} \label{3.18}
\eeqa
In these equations $Z_1(s)$ has accumulated the term with $n=0$
from Eq.\ (\ref{3.11}) subtracted by (\ref{3.14});
$Z_2(s)$ involves the contribution of the term of order $1/n^2$ in
expansion (\ref{3.13}) and the added expression (\ref{3.14});
$Z_3(s)$ is generated by the terms of order $1/n^4$ in the
expansion (\ref{3.13}).

Taking into account that
\beq\label{3.19}
\mu_0^2(y)\vert_{y\to 0}\to 1\quad
\mbox{and}\quad \mu_0^2(y)\vert_{y\to \infty}\to
\frac{1}{4y^2}+\frac{3}{16y^4}\,{,}
\eeq
the integration by parts in Eq.\ (\ref{3.16}) can be done  for
$-3<\mbox{Re}~s<1$ with the result
\beq\label{3.20}
Z_1(s)=\frac{s-1}{2\sqrt{\pi}\,a
\Gamma\left(\displaystyle\frac{s}{2}\right)
\Gamma\left(\displaystyle\frac{3-s}{2}\right)}
\left(\frac{c}{a}\right)^{-s}\int_0^{\infty}dy\,y^{-s}
\left\{\ln [1-\xi^2\mu_0^2(y)]+
\frac{\xi^2}{4}\,y^4t^6(y)\right\}\,{.}
\eeq
With allowance for (\ref{3.19}) one infer easily that the function $Z_1(s)$ is
an analytic function of the complex variable $s$ in
the region $-3<\mbox{Re}~s<1$.
In the linear order of $\xi^2$ it reduces to
\beq\label{Z1lin}
Z_1^{lin}(s)=\xi^2\displaystyle\frac{s-1}{2\sqrt{\pi}a
\Gamma\left(\displaystyle\frac{s}{2}\right)\Gamma
\left(\displaystyle\frac{3-s}{2}\right)}\left(\frac{c}{s}\right)^{-s}
\int_0^{\infty}dy\,y^{-s}\left[\frac{y^2}{4(1+y^2)^3}-\mu_0^2(y)\right]\,{.}
\eeq
This function is also analytic in the region $-3<\mbox{Re}~s<1$.
Integration in Eq.\ (\ref{3.17}) can be accomplished exactly by
making use of the formula
\beq\label{3.21}
\int_0^{\infty}dz\,z^{1-s}\frac{d}{dz}(z^4t^6)=\frac{s-1}{4}
\Gamma\left(\displaystyle\frac{1+s}{2}\right)
\Gamma\left(\displaystyle\frac{5-s}{2}\right)\,{,}
-1<\mbox{Re}~s<5\,{.}
\eeq
This gives for $Z_2(s)$ in (\ref{3.17})
\beq\label{Z2}
Z_2(s)=\xi^2\left(\frac{c}{a}\right)^{-s}
\frac{(1-s)(3-s)}{64\sqrt{\pi}\,a}\left(
2\sum_{n=1}^{\infty}n^{-s-1}+1\right)
\frac{\Gamma\left(\displaystyle\frac{1+s}{2}\right)}{
\Gamma\left(\displaystyle\frac{s}{2}\right)}\,{.}
\eeq
In view of the sum over $n$ in (\ref{Z2}) the function $Z_2(s)$
is defined only for $s>0$.

For simplicity we apply in Eq.\
(\ref{3.18}) the integration by parts
which is correct for $-3<\mbox{Re}~s<2$ and leads to
the result
\beq\label{Z3}
Z_3(s)=\xi^2\left(\frac{c}{a}\right)^{-s}\frac{(1-s)(3-
s)(7s^2-4s-27)}{6144\sqrt{\pi}\,a}
\frac{\Gamma\left(\displaystyle\frac{3+s}{2}\right)}{
\Gamma\left(\displaystyle\frac{s}{2}\right)}\,\sum_{n=1}^{\infty}n^{-s-3}\,{.}
\eeq
Again the sum over $n$ in (\ref{Z3}) gives the restriction
Re~$s>-2$ for definition of the function $Z_3(s)$.

Thus the spectral zeta function $\zeta_{cyl}(s)$ in the linear approximation
with respect to $\xi^2$ and with allowance for the first two terms in the
UAE (\ref{3.13}) is given by
\beq\label{3.25}
\zeta_{cyl}(s)=Z_1^{lin}(s)+Z_2(s)+Z_3(s)\,{,}
\eeq
where the $Z$'s are presented in Eqs. (\ref{Z1lin}), (\ref{Z2}) and
(\ref{Z3}), respectively.
Summing up all the restrictions on the
complex variable $s$ which have been imposed
when deriving Eqs.
(\ref{Z1lin}), (\ref{Z2}), and (\ref{Z3}),
we infer that $\zeta_{cyl}(s)$ is defined
in the strip $0<\mbox{Re}~s<1$. In order to continue
these equations into the surroundings
of the point $s=-1$,
it is sufficient to express the sum in Eq.\ (\ref{Z3})
in terms of the Riemann $\zeta$ function
\beq\label{Z3con}
Z_2(s) =\xi^2\left(\frac{c}{a}\right)^{-s}\frac{(1-s)(3-s)}{64\sqrt{\pi}\,
a}\,[2\zeta (s+1)+1]
\displaystyle\frac{\Gamma\left(\displaystyle\frac{1+s}{2}\right)}
{\Gamma\left(\displaystyle\frac{s}{2}\right)}\,{.}
\eeq
It is left now to take the limit $s\to -1$ in Eqs.\ (\ref{Z1lin}),
(\ref{Z2}) and (\ref{Z3con}).
A special care should be paid when calculating this limit
in (\ref{Z3con}) in view of the poles of the function $\Gamma((1+s)/2)$
at this point. Using the values
\beq\label{3.27}
\zeta (0)=-\frac{1}{2}, \quad \zeta^{\prime}(0)=-\frac{1}{2}\ln 2\pi\,{,}
\quad \Gamma^{\prime}(1)=\gamma \,{,}
\eeq
one derives
\beq\label{3.28}
\lim_{s\to -1}[2\zeta (1+s)+1]\Gamma\left(\frac{1+s}{2}\right)=
\eeq
$$
=\lim_{s\to -1}[2\zeta (0)+2\zeta^{\prime}(0)(1+s)+
O((1+s)^2)+1]\left[\frac{2}{1+s}+
\gamma+O(1+s)\right]=-2\ln (2\pi )\,{.}
$$
With allowance for this we obtain from (\ref{Z3con})
\beq\label{Z2fin}
Z_2(-1)=\frac{c\xi^2}{2\pi a^2}\,\frac{1}{4}\ln (2\pi )\,{.}
\eeq
The appearance of the finite term proportional to
$\ln (2\pi )$ is remarkable for the problem under consideration.
It is derived here in a consistent way by making use of an analytic
continuation of the relevant spectral zeta function. In Ref.~\cite{MNN} it
was obtained in a more transparent though not rigorous way.

Gathering together Eqs. (\ref{Z1lin}), (\ref{Z3}) with $s=-1$ and
Eq. (\ref{Z2fin}) we have
\beqa
\zeta_{cyl}(-1) & = & \frac{c\xi^2}{2\pi a^2}\left\{\int_0^{\infty}
yd_y\left[\frac{y^4}{4(1+y^2)^3}-\mu_0^2(y)\right]+\frac{1}{48}
\sum_{n=1}^{+\infty}\frac{1}{n^2}+\frac{1}{4}\ln (2\pi )\right\}
\nonumber \\
                & = & \frac{c\xi^2}{2\pi
                a^2}(-0.490878+0.034269+0.459469)\nonumber \\
                & = & \frac{c\xi^2}{2\pi a^2}\, 0.002860\,{.}
                \label{cyl3}
\eeqa
This result is not the final answer in the problem in hand. The
point is that in view of severe cancellations in (\ref{cyl3}) the
contribution of the next term in the UAE (\ref{3.13}) proves to be
essential. Its account gives \cite{MNN}
\beq\label{3.31}
\zeta_{cyl}(-1)=0\,{.}
\eeq
Thus the Casimir energy of a compact cylinder possessing the same
speed of light inside and outside proves to be zero. The
consideration presented in this Section can be extended to the
next term of order $\sim 1/n^6$ in the UAE (\ref{3.13}) in a
straightforward way. Therefore we shall not present here these
rather cumbersome expressions~\cite{foot3}.

Now we address to the consideration of a special case when $\xi
=1$. It corresponds to a perfectly conducting cylindrical shell
\cite{MNN}. Instead of the expansion (\ref{3.13}) we have
\beq\label{3.32}
\ln\left\{1-\left[z\frac{d}{dz}(I_n(nz)K_n(nz))\right]^2\right\}=
\eeq
$$
=
-\frac{z^4t^6}{4n^2}\left[1+\frac{t^2}{4n^2}\left(3-
30t^2+35t^4+\frac{1}{2}\,z^4t^4
\right)+O(n^{-4})\right]\,{.}
$$
Proceeding in the same way as above we obtain for the spectral zeta
function concerned
\beq\label{zetasc}
\zeta_{cyl}^{shell}(s)=Z_1(s)+Z_2(s)+Z_3(s)\,{,}
\eeq
where $Z_1(s)$ is given by Eq. (\ref{3.20}) with $\xi =1$,
$Z_2(s)$ is the same as in Eq. (\ref{Z3con}), and $Z_3(s)$ now is
\beq\label{Z3N}
Z_3(s)=\frac{(1-s)(3-s)(71s^2-52s-235)}{61440\sqrt{\pi}a^{1-s}}\,\frac{
\displaystyle \Gamma\left(\frac{3+s}{2}\right)}{\displaystyle
\Gamma\left(\frac{s}{2}\right)}
\sum_{n=1}^{+\infty}n^{-3-s}\,{.}
\eeq
At the point $s=-1$ it has the value
\beqa
\zeta_{cyl}^{shell}(-1)&=&\frac{1}{2\pi a^2}\,(-0.6517)+
\frac{1}{2\pi a^2}\frac{7}{480}
\sum_{n=1}^{+\infty}\frac{1}{n^2}+\frac{1}{8\pi a^2}\,\ln (2\pi )
\nonumber \\
                       &=&\frac{1}{2\pi
                       a^2}(-0.6517+0.0240+0.4595) \nonumber \\
                       &=& -\frac{1}{a^2}\, 0.0268\,{.} \label{zetasc1}
\eeqa
This exactly reproduce the contribution of the first two terms in
calculations of the Casimir energy for cylindrical shell in Ref.
\cite{MNN}. With higher accuracy this energy is given by
\cite{RM}
\beq\label{ECfin}
E_{cyl}^{shell}=-\frac{1}{a^2}\, 0.01356\,{.}
\eeq

In a recent paper \cite{GR1} the vacuum energy of a perfectly
conducting cylindrical surface has been calculated to much higher
accuracy by making use of another version of the zeta function
technique. By integrating over $dk_z$ directly in Eq.\ (\ref{zcyl1})
the authors reduced this problem to investigation of the zeta
function for circle, which has been considered earlier by
introducing the partial wave zeta functions for interior and
exterior region separately. In this respect our approach dealing
only with one spectral zeta function for given boundary conditions
proves to be more simple and straightforward.

\section{Conclusion}

The method for constructing the spectral zeta functions developed
here proceeds from the contour integral representation with a
subsequent employment of the uniform asymptotic expansions for the
Bessel functions. Upon an analytic continuation the zeta functions
prove to be presented as (infinite) series over the Riemann
$\zeta$ functions with rapidly decreasing terms (see, for example,
Eqs.\ (\ref{con}), (\ref{shell3})).

We didnot pursue here the goal of obtaining high accuracy when
calculating the Casimir energy. In fact we seek to present the
consideration in such a form that no manifest divergencies appear.
An obvious advantage of the regularization method in hand does not
need any renormalization.

By treating the boundary condition given on an infinite cylinder,
we have clearly demonstrated the importance of a consistent
analytic continuation of the relevant spectral zeta function, in contrast
to identifying simply the sum of the type
$\sum_{n=1}^{\infty}n^{-s}$ with Riemann $\zeta$ function, in
order to involve correctly the contributions to the Casimir
energy proportional to
$\ln (2\pi )$.

Consideration in this framework of the same configuration
of vacuum electromagnetic field  but
with different velocities of light inside and outside the
boundaries probably will demand the modification of the definition
of the spectral zeta functions for incorporating in a proper way
the contact terms important in this case~\cite{MNg,BM,BMM}.

\centerline{\bf Acknowledgments}

This work has been completed during the stay of one of the authors
(V.V.N.) at Salerno University. It is a pleasant duty for him to
thank Prof.\ G.~Scarpetta, Dr.\ G.~Lambiase, and Dr.\ A.~Feoli for
kind hospitality. The work was accomplished by partial financial
support of the Heisenberg--Landau Program and by the fund MURST
ex 60\% and ex 40\% DPR 382/80.


\begin{thebibliography}{99}
\bibitem{foot1} Interest in calculating the Casimir energy of a material ball
               (or spherical cavity)
                in an infinite medium
                is caused, to some extent,  by recent attempts to elucidate the mechanism of
                   sonoluminescence and, in particular, by
               Schwinger's idea to use here the dynamical Casimir effect \cite{Milton}.
\bibitem{foot2} If we are considering only outgoing
            waves it is natural to use the solutions of the Maxwell equations which
           are finite
          (or at least do not grow infinitely) in the future. Such solutions should
          be proportional to $\exp{(-i\omega t)} H_{\nu}^{(1)}(\omega r)$
          when $\omega$ lies
          in the lower half--plane. For the upper half--plane $\omega$, the solutions
          describing outgoing waves and finite in the future should contain the factor
          $\exp{(i\omega t)} H_{\nu}^{(2)}(\omega r)$.
\bibitem{foot3}  More precisely, in Ref.\
                 \cite{MNN} not only the next term in UAE (\ref{3.13}) has been
                 taken into account but also the first 5 terms in the sum in
                 (\ref{3.18}) were taken exactly instead of using their asymptotics
                 $n^{-3-s}$. This turns out to be essential for reaching the
                 required precision. However this point is not critical for our
                consideration because $Z_3(s)$ does not need analytic
                continuation.
\end{thebibliography}
\end{document}